\documentclass[prl,reprint,twocolumn,superscriptaddress,amsmath,amssymb,floatfix,longbibliography]{revtex4-2}
\usepackage{soul}
\usepackage{orcidlink}
\usepackage{graphicx}
\usepackage{epstopdf}
\usepackage[utf8]{inputenc}
\usepackage{amsmath,amsthm,amssymb}
\usepackage{latexsym}
\usepackage{bm}
\usepackage{dcolumn}
\usepackage{braket}
\usepackage[normalem]{ulem}
\usepackage{times}
\usepackage{hyperref}
\usepackage[section]{placeins}
\usepackage{float}
\usepackage{natbib}

\newcommand{\bq}{\mbox{\boldmath$q$}}
\newcommand{\bQ}{\mbox{\boldmath$Q$}}

\begin{document}

\date{\today}

\title{Phonon collapse reveals electronically driven CDW in UPt$_2$Si$_2$}
\title{Phonon collapse at the charge density wave transition in UPt$_2$Si$_2$}
\title{Anomalous phonon collapse induced by charge density wave in UPt$_2$Si$_2$}
\title{Phonon anomaly and breakdown induced by charge density wave in UPt$_2$Si$_2$}
\title{Large Kohn anomaly and phonon breakdown induced by charge density wave in UPt$_2$Si$_2$}
\title{Large Kohn anomaly and phonon collapse induced by charge density wave in UPt$_2$Si$_2$}

\author{Jooseop Lee}
\affiliation{Department of Science, Hongik University, Seoul, 04066 Republic of Korea}
\affiliation{CALDES, Institute for Basic Science, Pohang 37673, Republic of Korea}
\author{Greta L. Chappell}
\affiliation{Department of Physics, Florida State University, Tallahassee, Florida 32306, USA}
\affiliation{National High Magnetic Field Laboratory, Florida State University, Tallahassee, Florida 32310, USA}
\author{Ryan E. Baumbach\orcidlink{0000-0002-6314-3629}}
\thanks{Current affiliation: Department of Physics, University of California, Santa Cruz, California 95064, USA}
\affiliation{Department of Physics, Florida State University, Tallahassee, Florida 32306, USA}
\affiliation{National High Magnetic Field Laboratory, Florida State University, Tallahassee, Florida 32310, USA}
\author{Ayman H. Said}
\affiliation{Advanced Photon Source, Argonne National Laboratory, Lemont, Illinois 60439, USA}
\author{Igor A. Zaliznyak\orcidlink{0000-0002-8548-7924}} %
\email{zaliznyak@bnl.gov}
\affiliation{Condensed Matter Physics and Materials Science Division, Brookhaven National Laboratory, Upton, NY 11973, USA}

\begin{abstract}

Using high-energy-resolution inelastic x-ray scattering, we observe anomalous softening and damping of the transverse acoustic phonon in UPt$_2$Si$_2$ as the system is cooled towards the charge density wave (CDW) transition temperature, T$_{\rm{CDW}}$. The phonon exhibits a marked Kohn-type anomaly around the CDW wave vector, Q$_{\rm{CDW}}$, and becomes overdamped within a finite momentum range already well above T$_{\rm{CDW}}$. The dispersion anomaly is consistent with potential Fermi surface nesting, which together with the extended phonon collapse indicate strong electron-phonon coupling. The transition temperature estimated from the phonon softening is markedly lower than T$_{\rm{CDW}}$, consistent with a primarily electronic instability rather than a phonon-driven transition. Our results establish UPt$_2$Si$_2$ as a prime example of a strongly correlated electron CDW system with exceptionally strong electron-phonon coupling driving phonon softening and collapse.

\end{abstract}

\maketitle

A charge density wave (CDW) is a periodic modulation of the electronic charge density and an associated atomic displacements \cite{Gruner_book, GrunerRMP1988}. Understanding the nature of such macroscopic quantum phase is of fundamental importance, not least due to its proximity to unconventional superconductivity in the phase diagram of many layered materials \cite{Abbamonte2014, Hayden2012, KeimerNatPhys2014, Yazdani2014, Dean2019, Lee_etal_Abbamonte_PNAS2022, HaydenTranquada_2024}. Nevertheless, advancing such understanding proved to be a major challenge as CDWs appear in a puzzling variety and the underlying mechanisms are widely debated \cite{Morosan2016, Plummer2017, Rossnagel2011}. CDWs are often observed in low dimensional systems \cite{Mook1976, PougetHennion_etal_PRB1991, Jerome1979, Mou_PRB2014}; the concept of CDW was originally presented in the Peierls’ model of 1D chain composed of equally spaced atoms with a single unpaired electron \cite{Peierls_book}. Due to the perfect nesting of the Fermi surface by a single wave vector in a 1D system, the real part of the Lindhard electronic susceptibility diverges, which leads to a CDW associated with lattice dimerization and opening of a band gap at the Brillouin zone (BZ) boundary.

While this picture proved successful in explaining the CDW in several quasi-1D systems \cite{Mook1976, PougetHennion_etal_PRB1991, Jerome1979}, it became clear that in higher dimensions CDW can rarely be explained solely by the Fermi surface nesting (FSN) instability. It was noted that logarithmic divergence of the susceptibility can be fragile against small deviations from perfect nesting condition and therefore the $\bq$-dependent electron-phonon (e-ph) coupling may play an essential role in stabilizing CDW, such as in quasi-2D transition metal dichalcogenides \cite{Johannes_PRB2006, JohannesMazin_PRB2008, Weber_etal_PRL2011, Weber_etal_Reznik_PRL2011, Weber_PRB2015, Degiorgi2013}. In this respect, the CDW is viewed as a soft-phonon driven structural phase transition.

This view came to close scrutiny in the case of strongly correlated cuprates. Recent experiments find the CDW as a universal phenomenon across essentially all hole-doped cuprates \cite{Hayden2012, KeimerNatPhys2014, Yazdani2014, Dean2019, Lee_etal_Abbamonte_PNAS2022, HaydenTranquada_2024}, while the relation between CDW and superconductivity (SC) remains a subject of vigorous debate. The Hubbard model and its variants predict CDW as an intrinsic state of strongly correlated electrons in square lattices \cite{Chan2017, Devereaux2017, Gull2017}. The electron-electron (e-e) correlation is considered to be at the heart of CDW in cuprates \cite{Morosan2016, Plummer2017}. Moreover, pseudogap and superconducting phase are believed to be intimately related with a CDW \cite{DavisPNAS2019, LoretNatPhys2019}. In this case, lattice modulation is believed to be driven by e-e interaction and charge redistribution.

Identifying the mechanism of CDW experimentally is far from straightforward. In a real material, electrons and lattice are intrinsically coupled, e-ph coupling is often comparable to e-e interaction and their contributions are hard to disentangle \cite{Kim_NatComm2021, Rossnagel2011}. Time-resolved pump-probe measurements have been used to distinguish periodic lattice modulation from CDW based on different characteristic time-scales of electrons and phonons \cite{RossnagelNatComm2012,Meng2020,Otto_SciAdv2021}. In transition metal dichalcogenides, the role of $\bq$-dependent e-ph coupling was also investigated by comparing the observed phonon linewidth with the electronic joint density of states \cite{Weber_etal_PRL2011, Weber_etal_Reznik_PRL2011, Weber_PRB2015}. The conclusions from these approaches are, however, still limited \cite{Weber_etal_PRL2011,Meng2020,Otto_SciAdv2021}. 

Recently, a CDW was discovered in UPt$_2$Si$_2$ below T$_{\rm{CDW}}$ $\approx$ 320 K \cite{JLee2020}. The CDW involves strongly correlated U-5$\textit{f}$ states and coexists with an antiferromagnetic order below T$_{\rm{N}}$ = 35 K \cite{JLee_PRL2018,SteemanJPCM1990, Nieuwenhuys_PRB1987}.
%
Remarkably, the CDW modulation wave vector is close to the FSN wave vector of URu$_2$Si$_2$ \cite{BareilleNatComm2014, ElgazzarNatMat2009}, implying that similar Fermi surface driven physics might be at play. URu$_2$Si$_2$ is a heavy-fermion unconventional superconductor with the ``hidden order'' precursor phase whose origin is still debated \cite{MydoshOppeneer_RevModPhys2011,MydoshOppeneer_JPCM2020}.
The CDW wave vector of UPt$_2$Si$_2$ is consistent with the electronic band structure calculations \cite{JLee2020} and is clearly distinct from that of related 4$\textit{f}$ systems, RPt$_2$Si$_2$ (R: 4$\textit{f}$ rare earth), having $\approx$ 3 unit-cell period \cite{Ueda2013, Kim_SciRep2015}, but is very close to the $\approx 5$ unit-cell periodicity found in 5$\textit{f}$ URu$_2$Si$_2$. This suggests an important role of strongly correlated U-5$\textit{f}$ electrons in the CDW formation. Hence, elucidating the lattice response to the CDW in UPt$_2$Si$_2$ can provide important new insights into electronic mechanisms of CDW.

Here, we study the anomalous phonon behavior in UPt$_2$Si$_2$ near the CDW wave vector using high-energy-resolution inelastic x-ray scattering. We observe a marked softening of the transverse acoustic (TA) phonon mode and its strong damping on approaching T$_{\rm{CDW}}$, while there is no discernible change in the longitudinal acoustic (LA) phonon. The energy renormalization and linewidth broadening of the TA mode occur over a considerable range of the BZ, 0.3 $\lesssim$ H $\lesssim$ 0.5, indicating a giant Kohn anomaly \cite{KeimerNatPhys2014, PougetHennion_etal_PRB1991}. Furthermore, we find that the TA phonon is not just strongly anharmonic, but is already collapsed (over-damped) well above T$_{\rm{CDW}}$, which suggests that it is not a primary driver of the CDW transition. 

\begin{figure}[tp!]
\centering
\includegraphics[width=1.0\columnwidth]{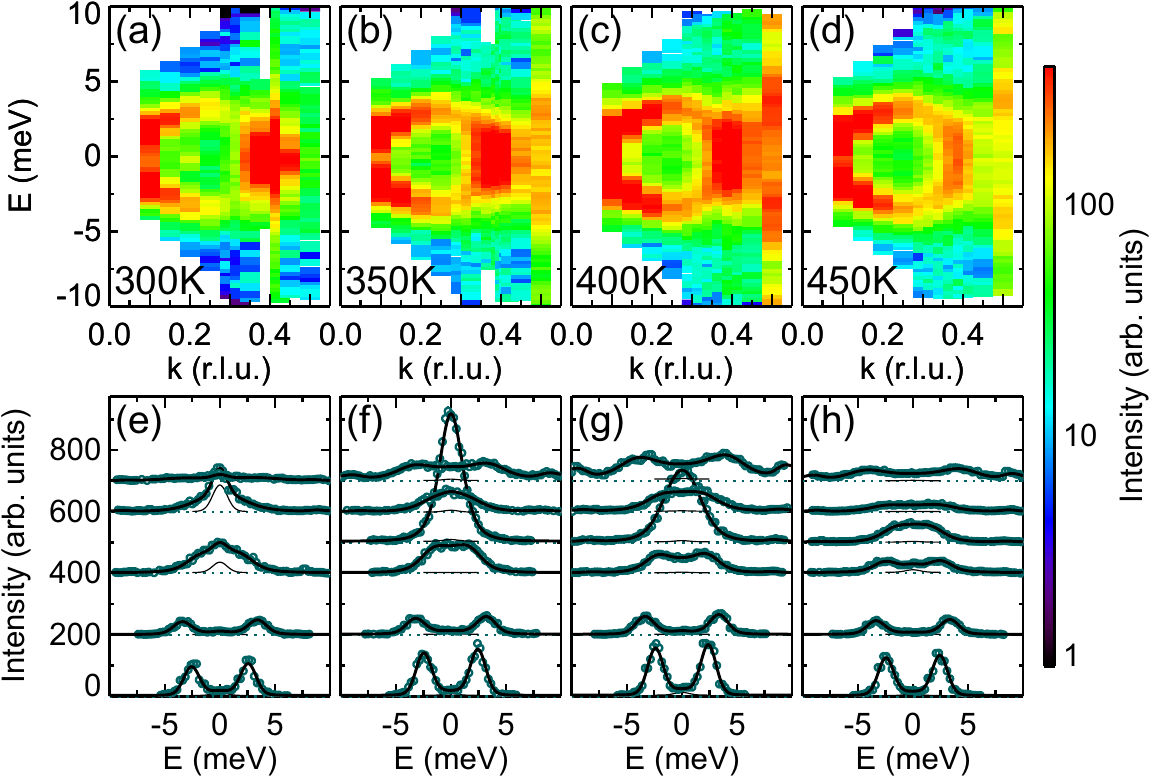}
\caption{The transverse phonon spectra at $\bQ$ = (4, k, 0) across T$_{\rm{CDW}}$ $\approx$ 320 K. (a)-(d) Color maps of the scattering intensity at T~=~300~K, 350~K, 400~K, and 450~K as a function of energy and momentum transfer, k, along the K-direction. (e)-(h) The line cuts of the data in panels (a)-(d), correspondingly, for k = 0.15, 0.25, 0.35, 0.4 (except at 300~K$< {\rm T_{CDW}}$), 0.45, and 0.5 (symbols) with fits to Eq.~\eqref{Eq:eq1} plus elastic peak at $E = 0$ (solid lines; thin lines are the fitted elastic peak). To present the intensity measured at different T on the same scale, the data in (e)-(h) was multiplied by 100/T. To avoid overlaps, the scans are shifted by $\Delta I \propto $~(k~-~0.15); the horizontal dashed lines indicate zero level for each scan. Error bars represent one standard deviation and where not visible are smaller than the symbol size. }
\label{fig:1}
\vspace{-0.2in}
\end{figure}

To explore the phonon dispersion curves of UPt$_2$Si$_2$, we used HERIX beamline at sector 30-ID-C of the Advanced Photon Source \cite{AymanJSR2020}. The highly monochromatic x-ray beam has an incident energy of 23.7 keV and provides an energy and wave vector resolution of $\approx$~1.5~meV and $\approx$~0.01~\AA$^{-1}$ full width at half maximum (FWHM), respectively \cite{Supplementary}. We index wave vectors in reciprocal lattice units (r.l.u.) of $P4/nmm$ crystal structure of UPt$_2$Si$_2$ with $a = b = 4.20$~\AA, $c = 9.63$~\AA. 

The transverse phonon spectra were measured by scanning energy at a constant wave vector transfer, $\bQ$ = (4, k, 0). The phonon scattering intensity prefactor, $(\bQ \cdot \bm{\epsilon})^2$, where $\bm{\epsilon}$ is the phonon polarization vector, ensures that only the transverse phonon modes are observed in these scans \cite{Shirane_book2002}. Fig.~\ref{fig:1} shows the intensity measured at several temperatures across the T$_{\rm{CDW}}$ $\approx$ 320 K. Even at the highest temperature, 450~K, the TA phonon is markedly softened and heavily damped around the CDW wave vector $\rm{\bQ_{CDW}} \approx$ (4, 0.4, 0). With decreasing temperature, the phonon gets broader and softer until most of its spectral weight condenses into an elastic peak with a broad quasi-elastic energy tail below T$_{\rm{CDW}}$ (see also Figs.~S4 and S5 \cite{Supplementary}). Already at 400 K the phonon mode near $\rm{\bQ_{CDW}}$ can hardly be resolved due to heavy damping.

For the quantitative analysis accounting for the observed phonon damping, we fit the data to a dynamical structure factor, S($\bq$, E), of a damped harmonic oscillator (DHO) \cite{PougetHennion_etal_PRB1991, SapkotaPRB2020},
%
%
\begin{equation}
    S(\bq, E) = \frac{I_{\bq} Z_{\bq}}{\pi\left(1-e^{-E/k_{B}T}\right)} \frac{\gamma_{\bq} E}{(E^2 - E_{\bq}^2 )^2 + (\gamma_{\bq} E)^2} ,
\label{Eq:eq1}
\end{equation}
%
convoluted with the measured Gaussian instrumental resolution function. Here, $\gamma_{\bq}$ is the damping parameter quantifying the inverse phonon lifetime (in the under-damped regime, $\gamma_{\bq}$ is the FWHM of a Lorentzian peak in the imaginary dynamical susceptibility), $E_{\bq} = \hbar \omega_{\bq}$ is the energy of an undamped phonon at wave vector $\bq$ (an oscillator with an undamped frequency $\omega_{\bq}$), $I_{\bq}$ is the integral intensity, and $Z_{\bq}$ is the normalization factor ensuring that $I_{\bq} = \int_{-\infty}^{\infty} S(\bq,E) dE$. In the under-damped regime, the damped phonon energy (DHO frequency $\widetilde{\omega}_{\bq}$) is given by, $\widetilde{E}_{\bq}$ = $\hbar \widetilde{\omega}_{\bq}$ = $\sqrt{E_{\bq}^2 - (\gamma_{\bq}/2)^2}$. For $E_{\bq} < \gamma_{\bq}/2$, in the over-damped regime, the imaginary damped phonon frequency shows that the atomic motion is a pure relaxation, without an oscillatory component, indicating collapse of phonon-like vibrational dynamics.

\begin{figure}[tp!]
\includegraphics[width=1.0\columnwidth]{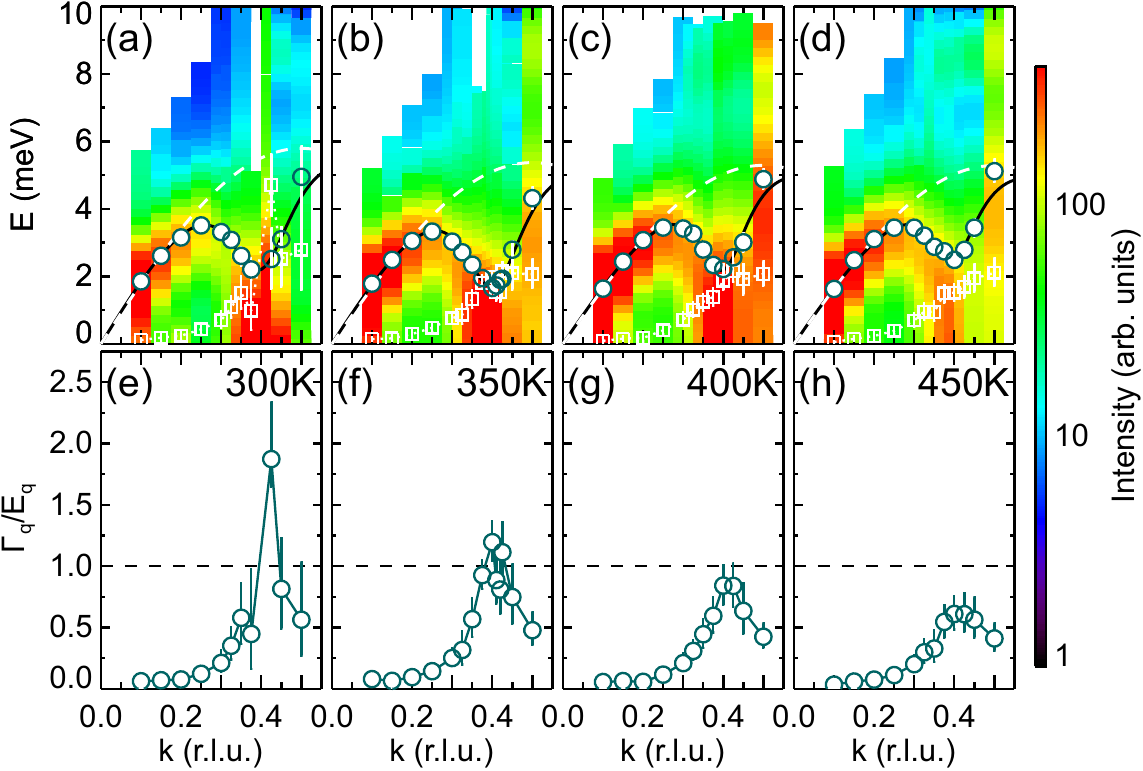}
\centering
\caption{The anomalous TA phonon dispersion and phonon collapse near $\rm{\bQ_{CDW}}$. (a)-(d) The circles (white-filled except where phonon within the error bar is over-damped) plotted over the color maps of the fitted intensity to the data in Fig.~\ref{fig:1} represent the fitted energy of the undamped phonon, $E_{q}$. The white squares connected with the broken white line show the DHO damping, $\Gamma_{\bq} = \gamma_{\bq}/2$ (Lorentzian HWHM in an under-damped case), proportional to the inverse lifetime of the phonon mode. The dashed white lines show fits to an acoustic phonon dispersion, $c|\sin(\pi k)|$ ($v_{TA} = \pi c$ is the TA sound velocity), in the limited range $k < 0.25$. The black solid lines show the full-range fits to the same phonon dispersion, but with a superimposed Lorentzian dip in energy phenomenologically describing a Kohn-like anomaly observed near $\rm{\bQ_{CDW}}$, at $k \approx$ 0.4 (see Supplementary \cite{Supplementary} for more details about the fits). (e)-(h) The ratio between $\Gamma_{\bq} = \gamma_{\bq}/2$ and $E_{\bq}$ shown in (a)-(d). An under-damped phonon exists for $\Gamma_{\bq}/E_{\bq}<1$ (horizontal dashed line). One standard deviation of the fitted values is smaller than the symbol size; the error bars shown correspond to a 50\% increase of the fit chi-squared. }
\label{fig:2}
\vspace{-0.2in}
\end{figure}

In all fits we also included a constant background and a resolution-limited Voigt peak with Gaussian FWHM of 1.5~meV at E = 0 to account for the possible elastic background, CDW peak below T$_{\rm{CDW}}$, and a potential central peak, a sharp quasi-elastic peak usually observed in conventional soft-phonon-driven structural phase transitions \cite{Hancock_PRB2015, Shapiro1972, CowleyShapiro2006}.

As shown in the bottom panel of Fig.~\ref{fig:1} and color maps in Fig.~\ref{fig:2}, the measured intensity is well-reproduced by the fits of the phonon spectra using Eq.~\ref{Eq:eq1} (where visible, we also add a small peak due to transverse optic phonon at $\approx$~9~meV). There is no feature in our data above T$_{\rm{CDW}}$ that can be attributed to a central peak (as shown by thin lines, the fitted central peak intensities are negligible and the fit quality is not improved by its addition). This is distinct from a soft-mode transition in a perovskite system where strong central peak at the superlattice position is observed in a broad wave vector and temperature range \cite{Hancock_PRB2015}, further corroborating a non-phonon-driven transition in our case (see Supplementary materials \cite{Supplementary} and Figs.~S4, S5, S13-S16 there for more details). 

In the top panels of Fig.~\ref{fig:2} we plot the fitted bare dispersion, $E_{\bq}$, over the fitted TA phonon spectra. For clarity, we focus on the positive energy transfer as the dynamical structure factor for the negative energy transfer is related by the principle of detailed balance, $S(\bq, -E) = e^{-E/k_{B}T} S(\bq, E)$. 
The white filled circles represent the bare, undamped phonon energy ($E_{\bq}$) and open squares show the damping parameter, $\gamma_{\bq}/2$. Fits to an acoustic phonon dispersion, $c|\sin(\pi k)|$ shown by dashed white lines reveal a marked dip in the measured TA dispersion near the CDW wave vector, at k~$\approx$~0.4. The dip is indicative of a large Kohn anomaly, which becomes more pronounced closer to T$_{\rm{CDW}}$. The black solid lines show fits to the same phonon dispersion but with a superimposed Lorentzian dip in energy phenomenologically describing the observed anomaly at $\rm{\bQ_{CDW}}$ (see Supplementary \cite{Supplementary} for fit details and results).
In addition to the Kohn-type anomalous decrease in the bare phonon (DHO) energy, we also observe a rapid increase of the phonon linewidth, $\gamma_{\bq}$, around $\rm{\bQ_{CDW}}$ as the sample is cooled down to T$_{\rm{CDW}}$, reflecting a broad, diffusive lattice response around the CDW wave vector.

The bottom panels of Fig.~\ref{fig:2} show the damping ratio, $\gamma_{\bq}/2E_{\bq}$, at each temperature, as a function of $\bq$. The ratio is strongly enhanced around $\rm{\bQ_{CDW}}$, it almost reaches 1 at 400~K and becomes $> 1$ already at 350 K, well above T$_{\rm{CDW}}$. When $\gamma_{\bq}/2E_{\bq}$ exceeds 1, the damped-phonon energy, $\widetilde{E}_{\bq}$, is no longer real and the atomic motion is a pure relaxation. The over-damped regime corresponds to the collapse of the TA phonon, that is, the poles are absent in the phonon Green’s function and no such quasiparticle exists, the phonon breaks down \cite{Nichitiu_PRB2024}.
The singular wave vector range where the phonon is over-damped increases as the temperature decreases towards T$_{\rm{CDW}}$. At 350~K the phonon is collapsed over a range of $\sim 0.1 \rm{\AA}^{-1}$. 

\begin{figure}[tp!]
\includegraphics[width=1.0\columnwidth]{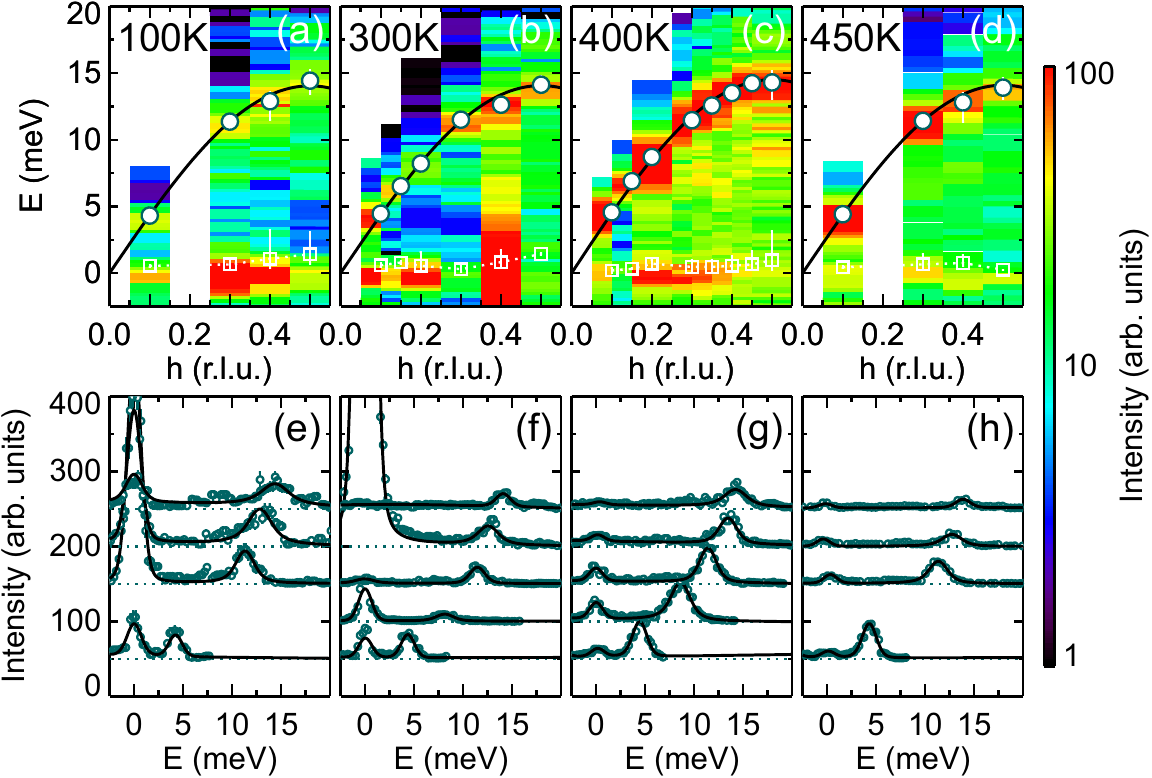}
\centering
\caption{The longitudinal phonon spectra around $\bQ$ = (4, 0, 0) across T$_{\rm{CDW}}$. (a)-(d) Color maps of the scattering intensity at T~=~100~K, 300~K, 400~K, and 450~K as a function of energy and momentum transfer, h, along the H-direction. The white-filled circles represent the fitted undamped phonon energy, $E_{\bq}$, and open white squares connected with dashed line the damping, $\gamma_{\bq}/2$. Solid lines are fits to a conventional acoustic phonon dispersion, $E_{\bq} = a|\sin(\pi h)|$. (e)-(h) The line cuts of the data in panels (a)-(d), correspondingly, (symbols) with fits to Eq.~\eqref{Eq:eq1} plus elastic peak at $E = 0$ (solid lines). To present intensity measured at different T on the same scale, the data in (e)-(h) was multiplied by 100/T. To avoid overlaps, the scans are consecutively shifted by $\Delta I \propto \Delta$~h; the horizontal dashed lines indicate zero level for each scan. Error bars in (a)-(d) correspond to a 50\% increase of the fit chi-squared, in (e)-(h) represent one standard deviation and where not visible are smaller than the symbol size. }
\label{fig:3}
\vspace{-0.2in}
\end{figure}

In a stark contrast to the TA phonon, the LA phonon mode shows no anomalous behavior. We measured the temperature and the wave vector dependence of the LA phonon by performing constant-$\bQ$ scans for $\bQ$ = (4+h, 0, 0). The top panels of Fig.~\ref{fig:3} show the measured x-ray scattering intensity with $E_{\bq}$ and $\gamma_{\bq}/2$ obtained by fitting to Eq.~\ref{Eq:eq1} plotted on top with white filled circles and open squares, respectively. The data and the fits are shown in the bottom panels of Fig.~\ref{fig:3}; they are in good agreement with each other.

Unlike for the TA mode, $\gamma_{\bq}$ is very small for the LA phonon (below the energy resolution) and does not show any measurable $\bq$- or temperature-dependence. The phonon is sharp, with negligible damping, and follows a sinusoidal dispersion curve expected for a conventional acoustic phonon, as shown by black solid line in the top panels of Fig.~\ref{fig:3}. We thus confirm that the LA mode does not play a role in the CDW formation and it is the TA phonon that is coupled with the CDW, consistent with the transverse atomic displacements observed in the previous x-ray diffraction experiment \cite{JLee2020}.

A soft-phonon-driven phase transition, as observed for example in perovskites, involves a well-defined phonon mode that softens at a particular wave vector, where a central peak typically emerges \cite{Hancock_PRB2015, Shapiro1972, CowleyShapiro2006}. In contrast, an overdamped lattice response indicates that a phonon eigenmode does not exist, and that lattice vibrations are not the relevant dynamical degrees of freedom. While lattice displacements remain important, the usual propagating phonon is absent, rendering the soft-mode picture inadequate in the regime of such extreme e-ph coupling. In our case, TA phonons are absent within a finite wave vector range around Q$_{\rm{CDW}}$, even well above T$_{\rm{CDW}}$. From the lattice-dynamical perspective, all these wave vectors are equivalent and characterized by diffusive behavior, with no phonon quasiparticles. This points to a mainly non-phonon-driven CDW mechanism, such as electron-electron or electron-lattice coupling, that both selects the particular ordering wave vector, $\mathbf{Q}_{\rm{CDW}}$, and overdamps the phonon.

\begin{figure}[tp!]
\includegraphics[width=1.0\columnwidth]{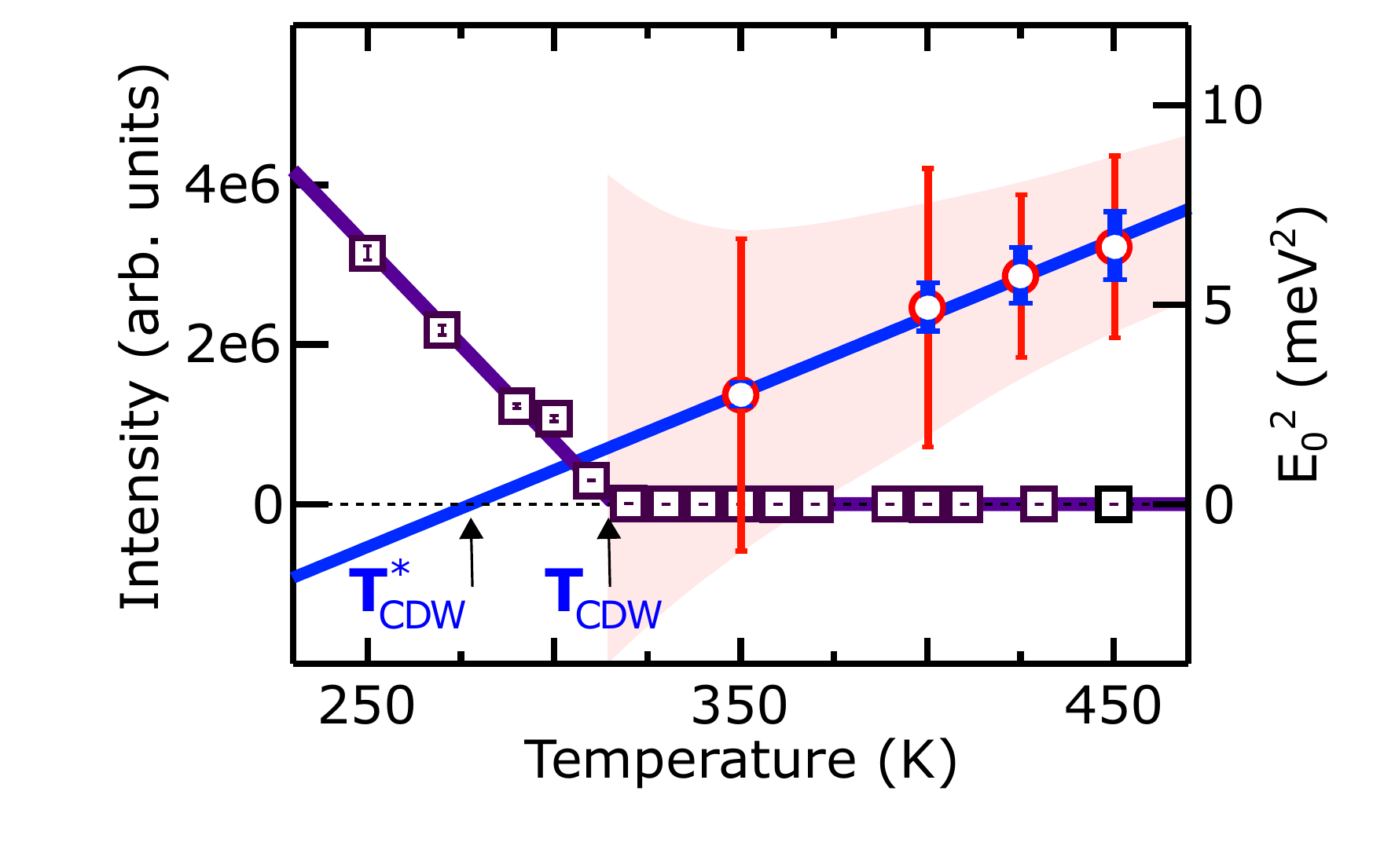}
\centering
\caption{The integrated elastic scattering intensity from a CDW satellite peak as a function of temperature (squares, left scale). The solid line is a fit to a mean-field behavior (for T $\ge$ 250 K) that gives a CDW onset temperature of T$_{\rm{CDW}} = 316(2)$~K. The error bars represent one standard deviation. Circles (right scale) are the (bare) un-damped energy squared of the TA phonon at $\rm{\bQ_{CDW}}$, which for the case of a soft-mode transition is expected to follow a linear temperature dependence above T$_{\rm{CDW}}^*$. In our case, linear fit (blue solid line) using standard deviations (thick blue error bars) gives T$_{\rm{CDW}}^* = 278(11)$~K. The vertical red bars whose magnitude increases precipitously below $\approx$~400~K correspond to $E_{\bq}^2 \pm (\gamma_{\bq}/2)^{2}$. Using these as errors in the linear fit yields T$_{\rm{CDW}}^* = 264(26)$~K.
}
\label{fig:4}
\vspace{-0.2in}
\end{figure}

The mechanisms underlying the CDW in UPt$_2$Si$_2$ are further revealed by Fig.~\ref{fig:4}. The square symbols show the temperature dependence of the integrated intensity of the elastic CDW peak, with solid line being a fit to the mean-field theory, $I \propto (T-T_{\rm{CDW}})$ \cite{Cowley_AdvPhys1980}. The obtained CDW transition temperature, T$_{\rm{CDW}} = 316(2)$~K, is consistent with the previous XRD measurement \cite{JLee2020}. The circles, on the other hand, show the bare (undamped) phonon energy squared at $\bq \approx \rm{\bQ_{CDW}}$, which roughly follows an $E_{\bq}^2 \propto (T-T_{c})$ relation obtained in the classical mean-field theory of soft-mode structural phase transitions \cite{Shirane_book2002,Cowley_AdvPhys1980,Cochran_AdvPhys1969,TkachenkoZaliznyak_PRB2021}. However, the transition temperature estimated from the extrapolated phonon softening (solid blue line), T$^*_{\rm{CDW}}$ = 278(11)~K, is noticeably lower than the actually observed onset temperature of the CDW.

Fig.~\ref{fig:4} suggests that an electronically driven soft-mode transition associated with the Kohn type phonon anomaly is preempted by a CDW instability that occurs at a higher temperature and is preceded by a TA phonon collapse near $\rm{\bQ_{CDW}}$ in a broad temperature range above T$_{\rm{CDW}}$. One might then argue that it is the energy cost of the collapse of lattice dynamics induced by strong e-ph coupling that favors CDW formation at T$_{\rm{CDW}} > $T$^*_{\rm{CDW}}$. This view is corroborated by an anomalous and unusual temperature dependence of the CDW in UPt$_2$Si$_2$, with an onset at a lattice-commensurate $\rm{\bQ_{CDW}} \approx$ (0.4, 0, 0) and a crossover to an incommensurate $\rm{\bQ_{CDW}} \approx$ (0.42, 0, 0) on cooling \cite{JLee2020}. The temperature where the wave vector locks into an incommensurate position is similar to T$^*_{\rm{CDW}}$ we observe here and is consistent with the onset temperature of the ``Kondo-lattice-type'' coherence, that is, a shift from local- to itinerant-electron behavior of U-5$f$ electrons \cite{JLee2020, Rosenbaum_NatPhys2015}.

It is illuminating to compare the CDW in UPt$_2$Si$_2$ with the electronic magnetic response in URu$_2$Si$_2$, which shows softening of magnetic excitation at $\rm{\bQ_{inc}}$ = (0.4 0 0), away from the magnetic zone center \cite{Wiebe_NatPhys2007}. The characteristic wave vector of electronic magnetism in URu$_2$Si$_2$ is determined by the nesting of the Fermi surface. This wave vector is similar to the $\rm{\bQ_{CDW}}$ $\approx$ (0.4 0 0) in UPt$_2$Si$_2$, which further supports involvement of 5$f$-electrons and FSN in the CDW.
There is an increasing consensus that the dual nature, both local and itinerant, persists throughout the UT$_2$Si$_2$ (T: transition metal) family with varying degrees of locality and itinerancy \cite{JLee_PRL2018, Severing_PNAS2020}.

In summary, we observe a large, Kohn-anomaly-type softening of the TA phonon around $\rm{\bQ_{CDW}}$ in a moderately enhanced 5$\textit{f}$-electron heavy fermion CDW system UPt$_2$Si$_2$. We also observe a breakdown of the lattice dynamics near $\rm{\bQ_{CDW}}$ well above the CDW transition temperature. The phonon collapse that we observe in metallic UPt$_2$Si$_2$ is distinct from the overdamped phonons at the Brillouin zone center observed in an insulating BaTiO$_3$ and relaxor ferroelectrics, where it is associated with the dynamics of polar nano-regions near the transition to ferroelectric phase \cite{CowleyShapiro2006}. Here, it occurs near the CDW wave vector consistent with the electronic instability governed by the nesting of the Fermi surface \cite{[{}] [{. While multiple nesting instabilities are predicted, the strongest is expected to be the most relevant for the electronic response.}] Elgazzar_PRB2012}. Furthermore, a pronounced Kohn anomaly around $\rm{\bQ_{CDW}}$ is observed in UPt$_2$Si$_2$ at temperatures as high as $\sim 1.5\cdot$T$_{\rm{CDW}}$. Overall, these results suggest a primarily electronic origin of the CDW in UPt$_2$Si$_2$, which together with the strong electron-lattice coupling have dramatic impact on the lattice dynamics.
Our results elucidate a rare case of correlated 5$\textit{f}$-electron system where CDW is primarily electronically driven.

\emph{Note added.}---After the initial submission of this manuscript, similar phonon softening and damping above T$_{\rm{CDW}}$, though less pronounced than in our case, were reported in the layered chalcogenide 2H-TaSe$_2$ \cite{Shen_NatComm2023}.

\begin{acknowledgments}

We thank Jason Jeffries for providing the sample cell. The work at Brookhaven National Laboratory was supported by Office of Basic Energy Sciences (BES), Division of Materials Sciences and Engineering,  U.S. Department of Energy (DOE),  under contract DE-SC0012704. This work was supported by IBS-R014-A2. This research used resources of the Advanced Photon Source, a US Department of Energy (DOE) Office of Science User Facility operated for the DOE Office of Science by Argonne National Laboratory under Contract No. DE-AC02-06CH11357.

\end{acknowledgments}


%


\begin{widetext}
\pagebreak
\end{widetext}
\hypersetup{pageanchor=false}
\renewcommand{\thepage}{S\arabic{page}}
\setcounter{page}{1}
\renewcommand{\theequation}{S\arabic{equation}}
\setcounter{equation}{0}
\renewcommand{\thefigure}{S\arabic{figure}}
\setcounter{figure}{0}
\renewcommand{\thetable}{S\arabic{table}}
\setcounter{table}{0}


\section*{Supplementary Information}

\begin{center}
{\bf Large Kohn anomaly and phonon collapse induced by charge density wave in UPt$_2$Si$_2$} \\
Jooseop Lee, Greta L. Chappell, Ryan E. Baumbach, Ayman H. Said, and Igor A. Zaliznyak 

correspondence to: zaliznyak@bnl.gov
\end{center}
\bigskip
\noindent{\bf This PDF file includes:}\\
Supplementary Text\\
Supplementary Tables S1--S3\\
Supplementary Figures S1--S16\\

\section{Sample preparation and characterization}
\label{sample}
\begin{figure}[hp]
\centering
\includegraphics[width=0.5\textwidth]{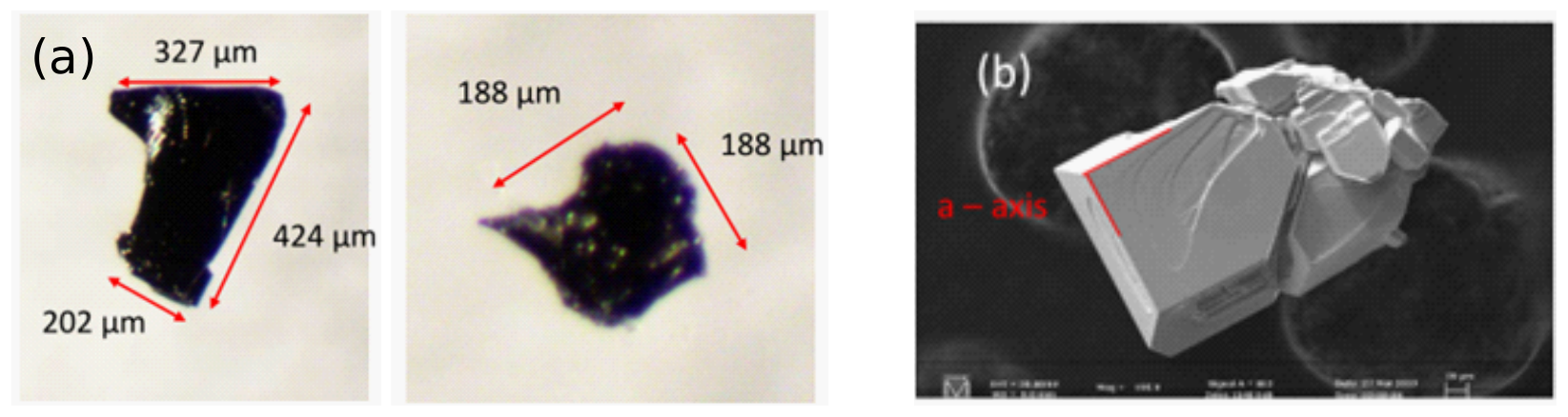}
\caption{(color online) (a) Image of UPt$_2$Si$_2$ crystals under the microscope demonstrating typical sample dimension of several hundred microns. (b) SEM image of an UPt$_2$Si$_2$ crystal where the red lines indicate the tetragonal ($P4/nmm$) $a$-axis. }
\label{fig:S1}
\vspace{-0.1in}
\end{figure}

Single crystals of UPt$_2$Si$_2$ were grown in an indium flux from elemental ingredients. The starting materials of 1(U):2(Pt):2(Si):40(In) with a purity higher than 99.9 $\%$ were sealed under argon atmosphere in a tantalum tube, which was then heated in a vertically translating induction furnace system. This process resulted in a high quality single-crystalline $ab$-plane platelets (Fig.~\ref{fig:S1}).

The stoichiometry and crystal structure were verified by energy-dispersive x-ray spectroscopy (EDS), transmission electron microscopy (TEM), and single crystal x-ray diffraction. The atomic percentages in chemical composition were determined as 19.45 $\%$, 41.04 $\%$, and 39.51 $\%$ for U, Pt, and Si, respectively.


\section{Inelastic x-ray scattering setup}
\label{experimental_setup}
\begin{figure}[hpt]
\centering
\includegraphics[width=0.5\textwidth]{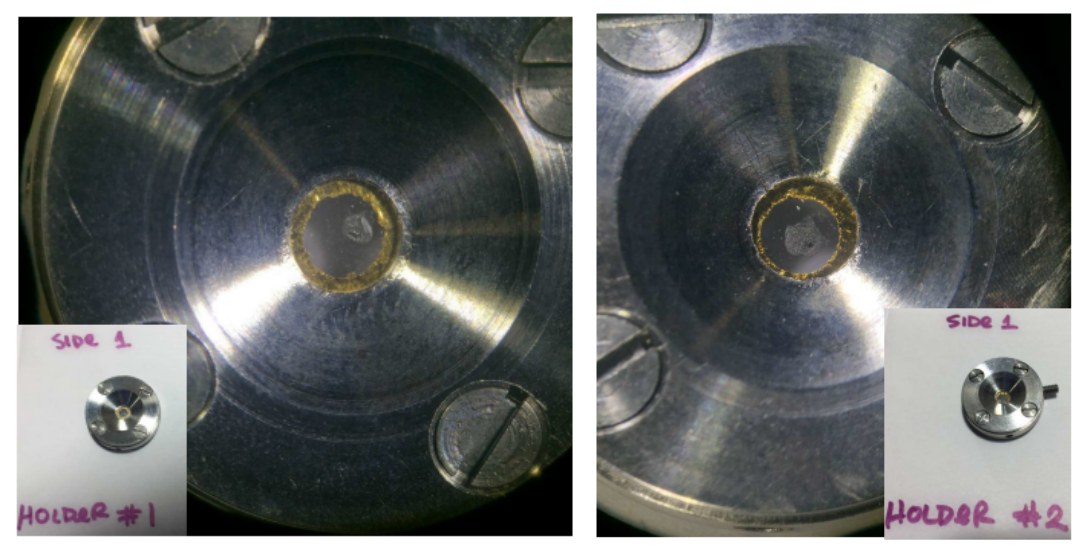}
\caption{(color online) Images of sample mounts used for inelastic x-ray scattering experiments on mildly radioactive UPt$_2$Si$_2$ single crystals. Platelet crystal samples are attached to a diamond plate with carbon paste, a gold foil gasket is placed over the plate, and another diamond plate is placed on top of the foil. The diamond plates are then sandwiched between the two holders that are held together with screws.}
\label{fig:S2}
\vspace{-0.1in}
\end{figure}

The monochromatic x-ray beam at HERIX has an incident energy of 23.7 keV and an energy resolution of $\approx$ 1.5 meV. A 15~mm diameter aperture in front of the analyzer provided the wave vector resolution of 0.0099 \AA$^{-1}$. The beam was focused on the sample to a cross-section of about 35 $\times$ 10 $\rm{\mu m^2}$ (horizontal $\times$ vertical). Small single crystals of area $\approx$ 250 $\times$ 250 $\rm{\mu m^2}$ and thickness of $\approx$ 10 $\rm{\mu m}$ (close to one absorption length at the given incident energy) were measured from T = 20 K to 450 K using a closed-cycle cryostat. The measurements were performed in the transmission geometry. For radiological containment, samples were sandwiched between diamond windows clamped with a steel holder. The samples were glued with a carbon paste and encapsulated with a gold foil gasket, Fig.~\ref{fig:S2}.

\begin{figure}[hpt]
\centering
\includegraphics[width=0.45\textwidth]{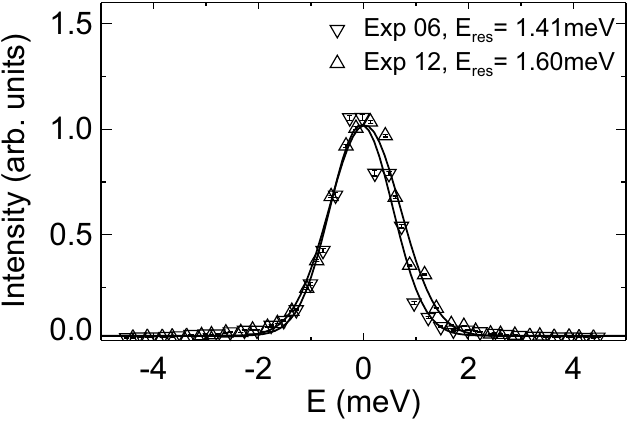}
\caption{(color online) Energy resolution of the HERIX setups used in our 06 and 12 experiments. The resolution FWHM is 1.5(1)meV.}
\label{fig:S3_HERIX_resolution}
\vspace{-0.1in}
\end{figure}

We have performed two consecutive measurements, tagged 06 and 12, in which a range of different temperatures were measured with an overlapping data set at T = 450~K. The same setup and the same sample were used in both experiments. Factor 1.38 accounts for the intensity cross-calibration between the data sets 06 and 12 at 450~K. The measured energy resolution of our setup is shown in Figure~\ref{fig:S3_HERIX_resolution}.

The longitudinal acoustic (LA) and transverse acoustic (TA) phonon dispersions were measured along the high-symmetry directions, (4 + h, 0, 0) and (4, k, 0), respectively. The measured spectra were fitted with a damped harmonic oscillator (DHO) lineshape (Eq.~1 in the main text) describing the phonon and a Lorentzian quasi-elastic peak, all convoluted with the Gaussian instrument energy resolution of $1.5$~meV FWHM measured at the elastic position (Fig.~\ref{fig:S3_HERIX_resolution}).

\section{Central peak}
\label{central_peak}
\begin{figure}[t!]
\centering
\includegraphics[width=0.5\textwidth]{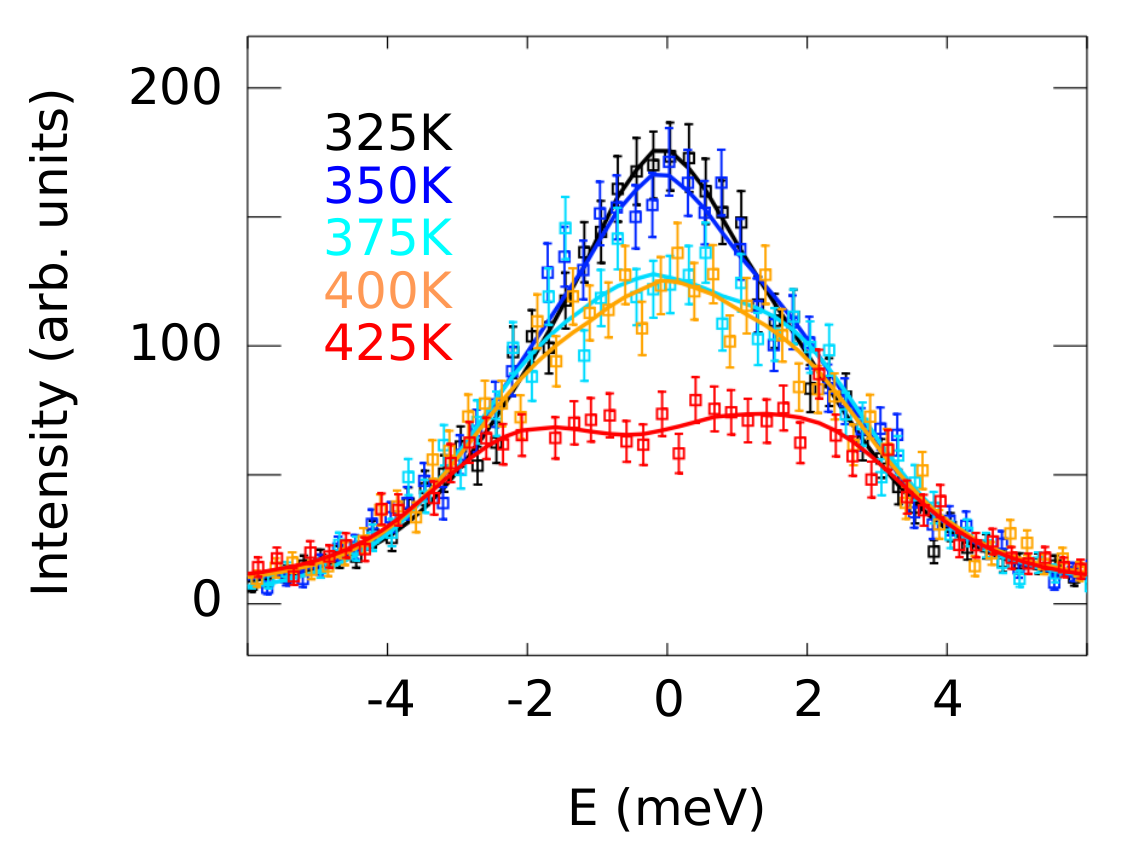}
\caption{(color online) Phonon energy spectra measured at various temperatures above T$_{\rm{CDW}}$ at a fixed $\bQ = (4, 0.45, 0)$. The error bars show one standard deviation. The solid lines are the DHO fits to Eq.~1 of the main text, where the fit results are shown in Figures 2 and 4 of the main text. }
\label{fig:S4}
\vspace{-0.1in}
\end{figure}

Figure~\ref{fig:S4} presents phonon energy spectra measured at various temperatures above T$_{\rm{CDW}}$ for a fixed $\bQ = (4, 0.45, 0)$. These line cuts are representative of the data shown in the color intensity maps in Figures 1 and 2 of the main text. The symbols represent the measured intensity and the solid lines show the fits to Eq.~1 of the main text, the phonon dynamical structure factor described by a damped harmonic oscillator. An excellent agreement of the data and the fit in Fig.~\ref{fig:S4} indicates the absence of any feature that can be attributed to a ``central peak''. An attempt to include a resolution limited elastic peak resulted in negligible fitted intensity. The same behavior is observed for other $\bQ$ positions (some are shown in Fig. 1 of the main text; for completeness, we include all data with fits below in this Supplementary).

\begin{figure}[h!]
\centering
\includegraphics[width=0.5\textwidth]{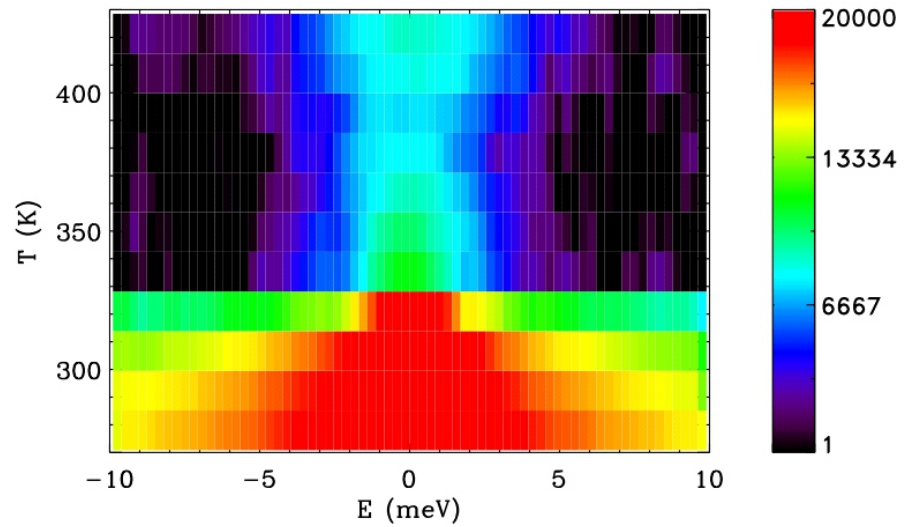}
\caption{(color online) The temperature dependence of a broad, [-10,10] meV energy scan at $\rm{\bQ_{CDW}} = (4.0,0.4,0)$, which illustrates the evolution of elastic and inelastic intensity at $\rm{\bQ_{CDW}}$.}
\label{fig:S5}
\vspace{-0.1in}
\end{figure}

The appearance of the elastic CDW peak intensity and the evolution of the inelastic, damped and over-damped-phonon intensity at $\rm{\bQ_{CDW}} = (4.0,0.4,0)$ further illustrating the absence of a central peak are presented in Figure~\ref{fig:S5}, which shows the temperature dependence of a broad, $[-10, 10]$~meV energy scan across the elastic position at $\rm{\bQ_{CDW}}$.

While their origin is still controversial, central peaks are a hallmark of soft-mode transitions that is observed in almost all systems with a phonon-softening-driven structural phase transition \cite{Hancock_PRB2015, Shapiro1972, CowleyShapiro2006, Cowley_AdvPhys1980, Fleury1976}. The absence of a central peak, therefore, can be considered as a strong circumstantial evidence supporting electronically driven nature of CDW in UPt$_2$Si$_2$.

\section{LA phonon dispersion fit}
\label{LA_phonon}
\begin{figure}[h!]
\centering
\includegraphics[width=0.5\textwidth]{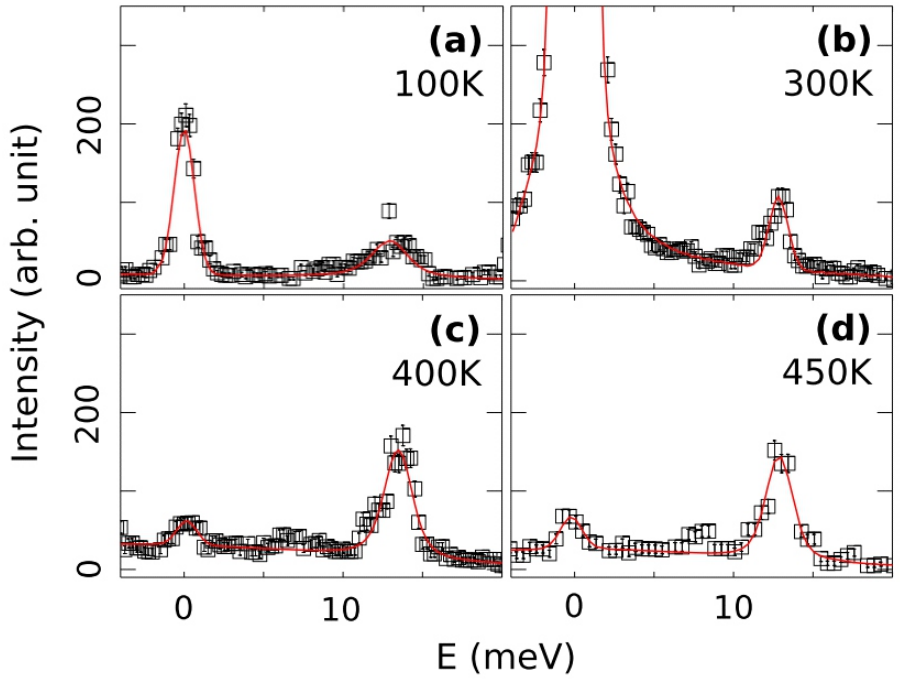}
\caption{(color online) The typical scans across the LA phonon at $Q = (4.4,0,0)$ at several temperatures. Symbols are the measured data, the error bars show one standard deviation. The solid lines are the fits to DHO (Eq.~1 of the main text) describing the LA phonon plus a quasi-elastic Lorentzian peak centered at $E = 0$, both convoluted with the Gaussian instrument energy resolution of $\approx 1.5$~meV.}
\label{fig:S6}
\vspace{-0.1in}
\end{figure}

For all temperatures, the LA phonon mode can be well described with a conventional acoustic phonon dispersion model, $\epsilon(h) = a~|\sin(\pi h)|$, where $h$ is the momentum transfer along the H direction in reciprocal lattice units (r.l.u.), and with the negligible damping, $\Gamma_{\bq}$, throughout the Brillouin zone. This is shown in Fig.~3 of the main text. The temperature evolution of a typical scan through the LA phonon mode, at $Q = (4.4, 0, 0)$, is presented in Figure~\ref{fig:S6}.

The fitted values of the phonon velocity (in meV/r.l.u.),  $v_{LA} = \pi a$, are summarized in Table~\ref{tab:S1}. With a high degree of accuracy, the LA phonon velocity is constant at all temperatures. This temperature-independence demonstrates surprising insensitivity of the lattice compressibility (bulk modulus) to the CDW transition, again consistent with the conduction-electron rather than lattice-vibrational origin of the CDW in UPt$_2$Si$_2$.

\begin{table}[h!]
\caption{\label{tab:S1} LA phonon velocity, $v_{LA} = \pi a$ (in meV/r.l.u.), obtained from the dispersion fits shown in Figure 3 of the main text.}
\begin{ruledtabular}
\begin{tabular}{lcccc}
\mbox{T, K}&\mbox{100 K}&\mbox{300 K}&\mbox{400 K}&\mbox{450 K} \\
\hline
$v_{LA} = \pi a$ & 44.1 $\pm$ 0.8 & 44.1 $\pm$ 1.3 & 45.5 $\pm$ 1.3 & 44.2 $\pm$ 0.8 \\
\end{tabular}
\end{ruledtabular}
\end{table}

\section{TA phonon dispersion fit}
\label{TA_phonon}
The observed Kohn-like anomalous softening of the TA phonon around $\b\rm{\bQ_{CDW}}$ cannot be described with the conventional acoustic phonon dispersion model, $\epsilon(k) = c~|\sin(\pi k)|$, where $k$ is the momentum transfer along K direction (in r. l. u.) and $v_{TA} = \pi c$ is the TA sound velocity (in meV/r.l.u.). This is clearly indicated by the fits shown by broken white lines in Fig.~2 of the main text. The fits were performed for $k <0.25$, where the effects of anomalous phonon softening are still insignificant. The TA phonon velocity, $v_{TA}$, obtained from these fits is presented in Table~\ref{tab:S2} and is roughly temperature-independent, once again indicating insensitivity of the bulk crystal lattice to the CDW transition.

In order to account for the observed Kohn anomaly of the TA phonon, we adopted a phenomenologicaly modified phonon dispersion with a Lorentzian dip at a wave vector $k_0$ describing the anomalous softening,
\begin{equation}
 \epsilon(k) = c~|\sin(\pi k)| \left[ 1-\frac{d}{1+\left( \frac{k-k_0}{\kappa}\right)^2} \right] .
\label{Eq:TA_disp}
\end{equation}
Here, in addition to $v_{TA}$ ($= \pi c$) the fitted parameters are the amplitude of the dip, $d$, its position, $k_0$, and the Lorentzian half-width at half-maximum (HWHM), $\kappa$. The fits are shown by the solid lines in Figure~2 of the main text and the fitted values of the parameters are listed in Table~\ref{tab:S3}. While parameters and the model are purely phenomenological, they have transparent physical meaning, specifying the position, the strength, and the width of the Kohn type anomaly in the TA phonon dispersion in UPt$_2$Si$_2$.
The amplitude of the dip is significant, $>$~50\%, and shows an increasing, softening trend with temperature decreasing towards T$_{\rm{CDW}}$. Notably, the refined dip position at $k_0 \approx 0.40$ remains temperature-independent, corroborating the electronic origin of the observed Kohn-like phonon anomaly. The TA phonon velocity obtained from the modified dispersion fits (Table \ref{tab:S3}) is in good agreement (within error) with the conventional dispersion fits in the limited k-range, $k <0.25$, presented in Table \ref{tab:S2}.

Below the CDW transition temperature, at 300~K, the TA phonon mode could be expected to become a soft Goldstone phonon, with zero energy gap at the CDW wave vector, $k_0$. However, for a range of wave vectors near $k_0$ the lattice response is overdamped and thus devoid of a wave-like oscillatory behavior, indicating still an extended region of phonon collapse at this temperature (Fig.~2(a),(e) of the main text). We also note that below T$_{\rm{CDW}}$ the overwhelming intensity of the elastic CDW peak makes it impossible to resolve phonons in the close, $\sim \pm 0.01$~r.~l.~u. vicinity of $\rm{\bQ_{CDW}}$.

\begin{table}[h!]
\caption{\label{tab:S2} TA phonon velocity, $v_{TA} = \pi c$ (in meV/r.l.u.), obtained from fitting the dispersion shown in Figure 2 of the main text to an acoustic phonon dispersion, $\epsilon(k) = c~|\sin(\pi k)|$, in the $k <0.25$ range where effects of anomalous phonon softening are insignificant (dashed white line in Fig.~2).}
\begin{ruledtabular}
\begin{tabular}{lcccc}
\mbox{T, K}&\mbox{300 K}&\mbox{350 K}&\mbox{400 K}&\mbox{450 K} \\
\hline
$v_{TA} = \pi c$ & 18.20 $\pm$ 0.92 & 16.87 $\pm$ 0.71 & 16.62 $\pm$ 0.19 & 16.60 $\pm$ 0.19 \\
\end{tabular}
\end{ruledtabular}
\end{table}

\begin{table}[h!]
\caption{\label{tab:S3} The fitted parameters of the modified TA phonon dispersion, Eq.~\eqref{Eq:TA_disp}, shown by the solid lines in Figure~2 of the main text: TA phonon velocity, $v_{TA} = \pi c$ (meV/r. l. u.), dip magnitude, $d$, position, $k_0$ (r. l. u.), and HWHM, $\kappa$ (r. l. u.).}
\begin{ruledtabular}
\begin{tabular}{lcccc}
\mbox{T, K}&\mbox{300 K}&\mbox{350 K}&\mbox{400 K}&\mbox{450 K} \\
\hline
$v_{TA}$    & 20.19 $\pm$ 0.73 & 19.18 $\pm$ 1.452 & 17.75 $\pm$ 1.02 & 17.32 $\pm$ 0.80 \\
$d$                 & 0.64 $\pm$ 0.04 & 0.70 $\pm$ 0.02 & 0.59 $\pm$ 0.04 & 0.52 $\pm$ 0.05 \\
$k_0$               & 0.394 $\pm$ 0.008 & 0.399 $\pm$ 0.005 & 0.397 $\pm$ 0.007 & 0.395 $\pm$ 0.012 \\
$\kappa$            & 0.106 $\pm$ 0.018 & 0.102 $\pm$ 0.0198 & 0.079 $\pm$ 0.017 & 0.079 $\pm$ 0.023 \\
\end{tabular}
\end{ruledtabular}
\end{table}

\section{Additional presentation of the data and the fits}
\label{Additional_data}

For completeness, we include in this Supplementary additional presentation of our data and fits which might be useful for evaluating our results.

\begin{figure}[h!]
\includegraphics[width=0.5\textwidth]{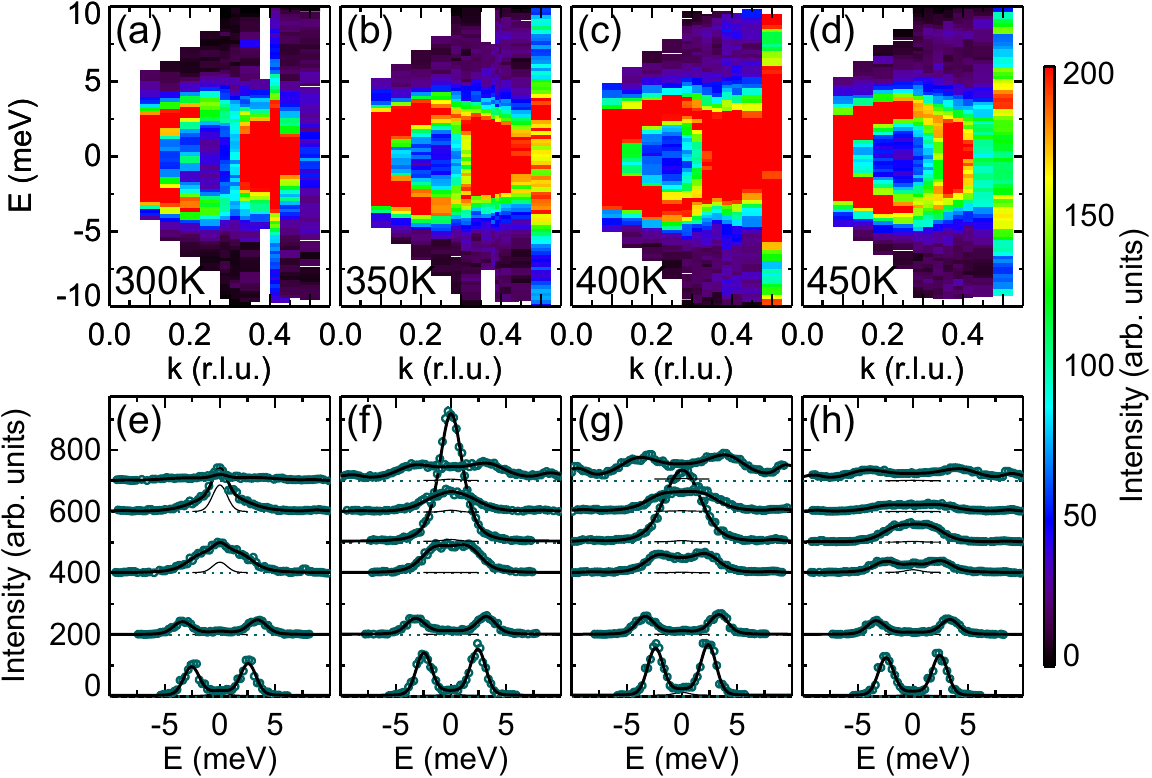}
\caption{(color online) A version of Figure~1 of the main text using a different intensity color map, where the same colors represent intensity rather than its logarithm as in Figure~1.}
\label{figS:FigA_new_linear_v1}
\vspace{-0.1in}
\end{figure}

\begin{figure}[h!]
\includegraphics[width=0.5\textwidth]{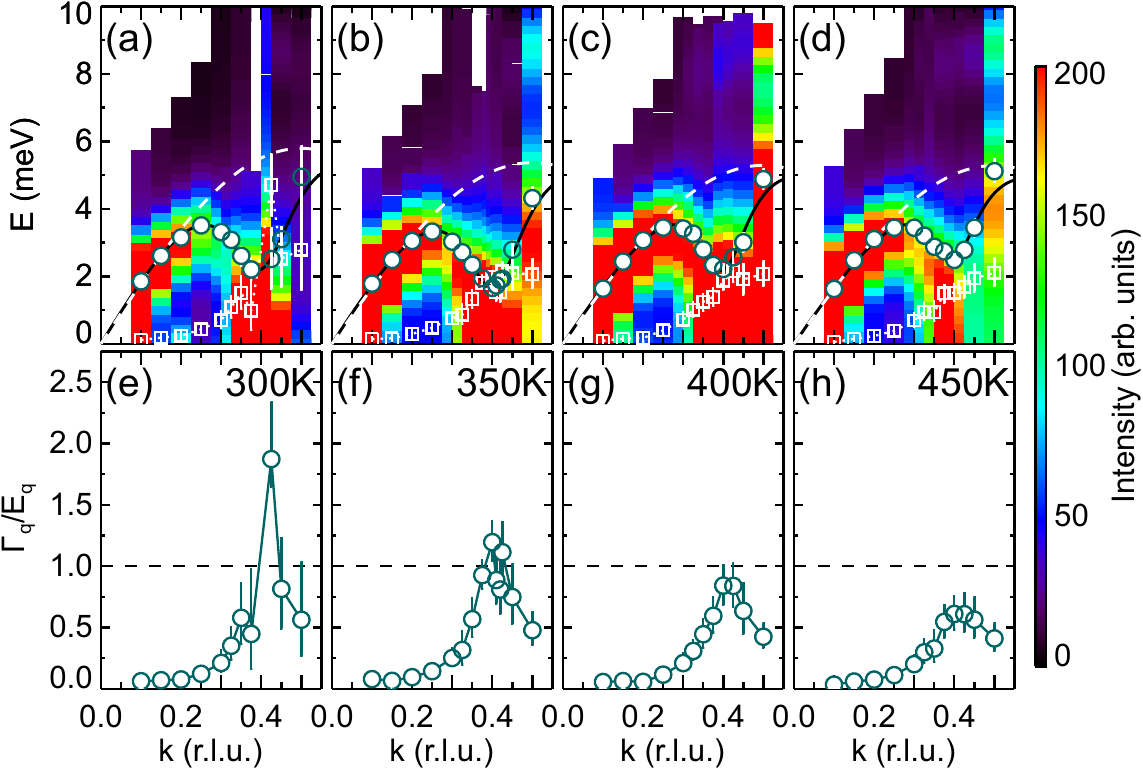}
\caption{(color online) A version of Figure~2 of the main text using a different intensity color map, where the same colors represent intensity rather than its logarithm as in Figure~2.}
\label{figS:FigB_new_linear_v1}
\vspace{-0.1in}
\end{figure}

Figure~\ref{figS:FigA_new_linear_v1} is a version of Figure~1 in the main text but using a different intensity color map, where the same colors are representing intensity rather than its logarithm as in Figure~1. Similarly, Figure~\ref{figS:FigB_new_linear_v1} is a version of Figure~2 in the main text using intensity color map where the same colors are representing intensity rather than its logarithm as in Figure~2. All important features of the data are observed independent of which color map representation is used.

\begin{figure}[h!]
\includegraphics[width=0.5\textwidth]{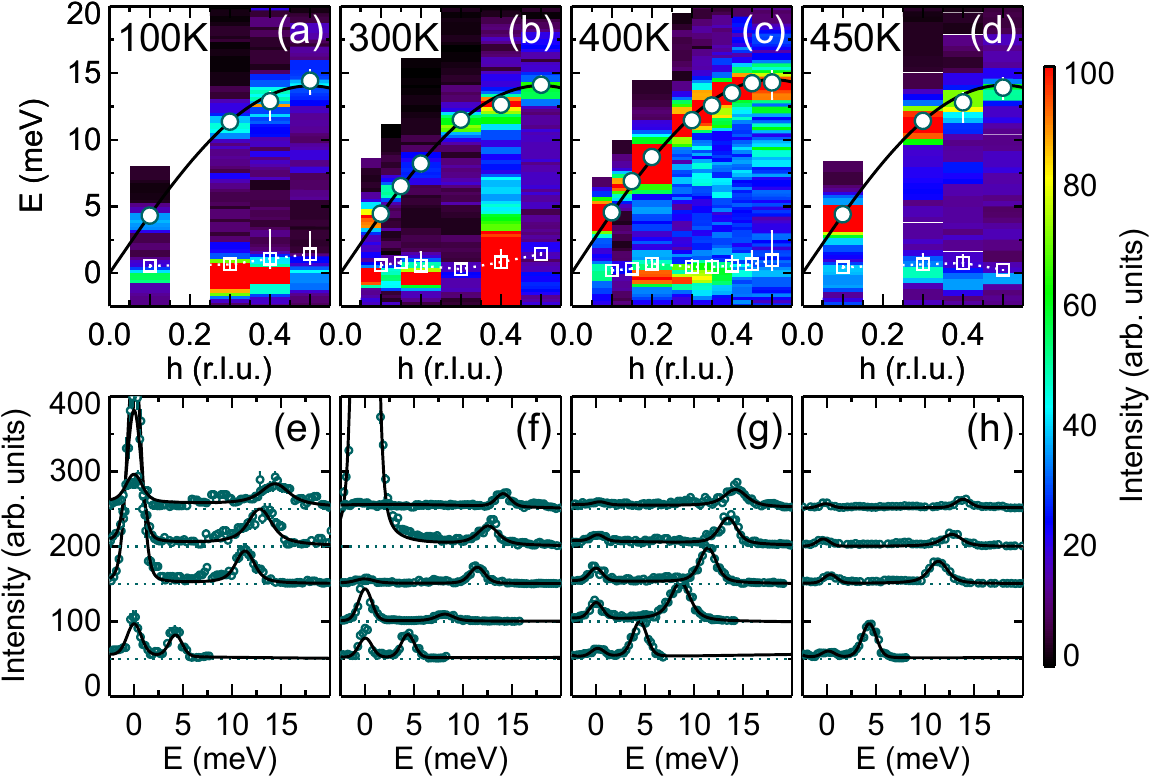}
\caption{(color online) A version of Figure~3 of the main text showing the LA phonon spectra and fits using a different intensity color map, where the same colors represent intensity rather than its logarithm as in Figure~3.}
\label{figS:FigC_new_linear_v1}
\vspace{-0.1in}
\end{figure}

Figure~\ref{figS:FigC_new_linear_v1} is a version of Figure~3 in the main text using intensity color map where the same colors are representing intensity rather than its logarithm as in Figure~3. All important features of the data are observed independent of which color map is used.

\begin{figure}[h!]
\includegraphics[width=0.5\textwidth]{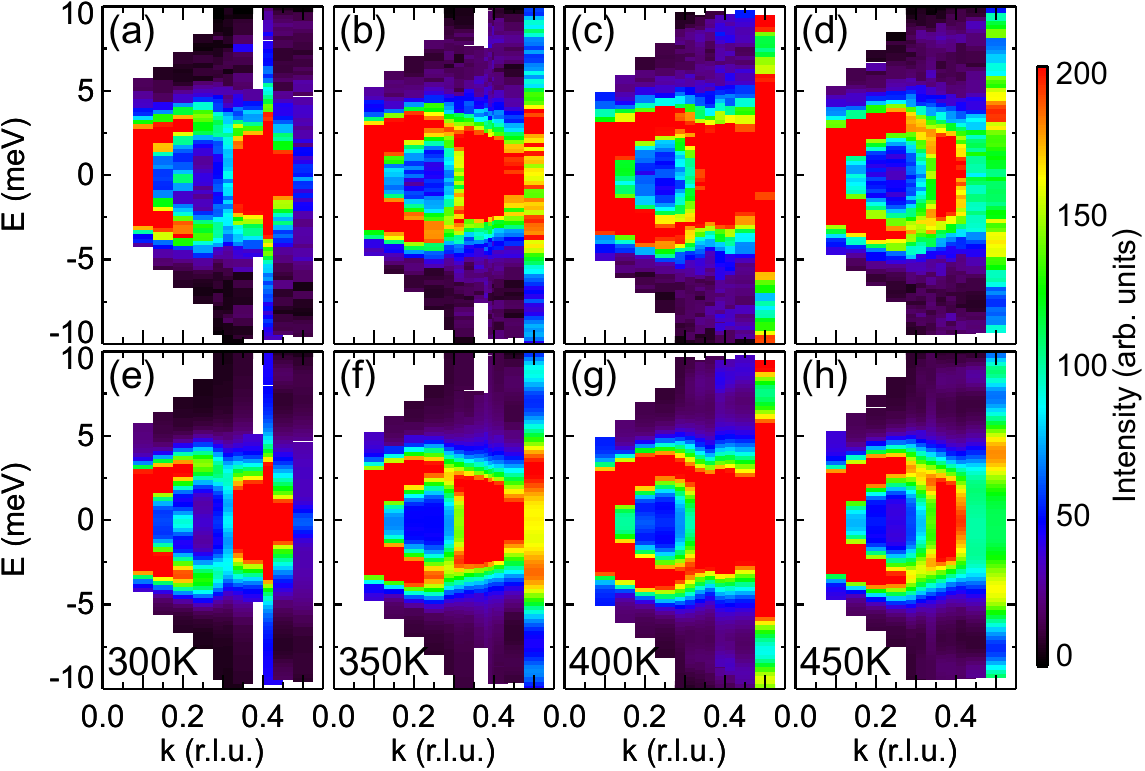}
\caption{(color online) The transverse phonon spectra at $\bQ$ = (4, k, 0) across T$_{\rm{CDW}}$ $\approx$ 320 K. (a)-(d) Color maps of the scattering intensity at T~=~300~K, 350~K, 400~K, and 450~K as a function of energy and momentum transfer, k, along the K-direction. The bottom panels, (e)-(h), show the fitted intensity using Eq.~(1) of the main text.}
\label{figS:FigA_Supplementary_linear}
\vspace{-0.1in}
\end{figure}

\begin{figure}[h!]
\includegraphics[width=0.5\textwidth]{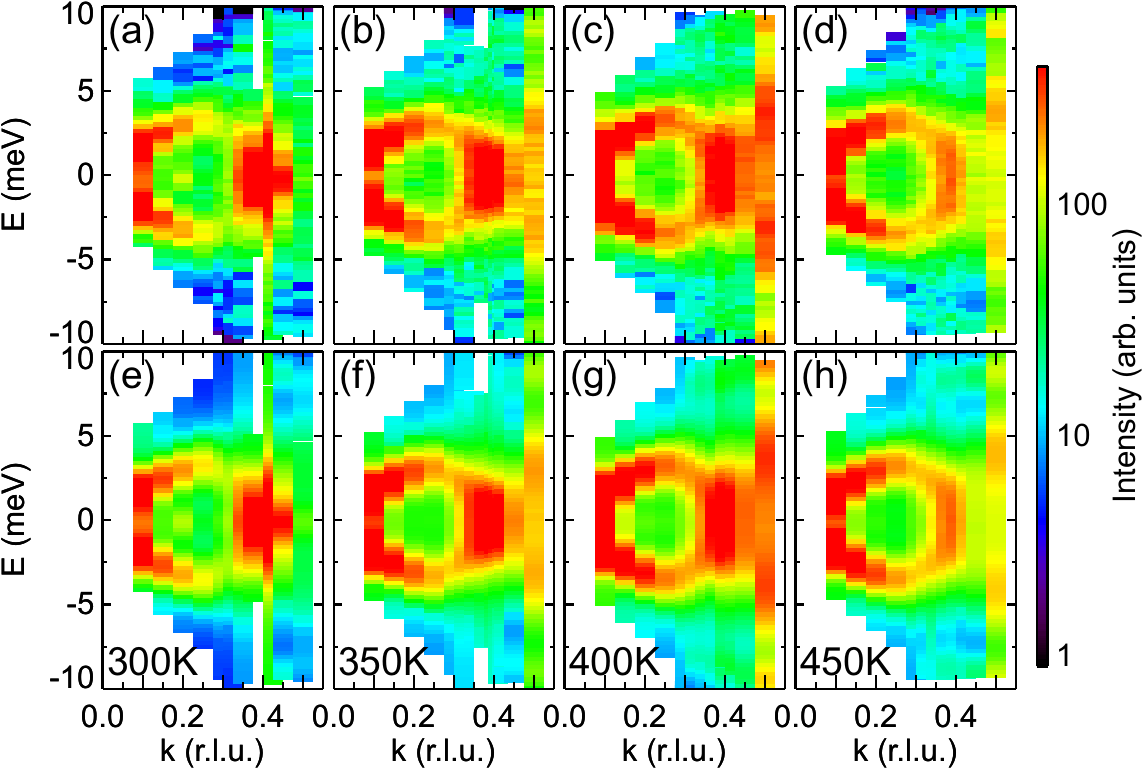}
\caption{(color online) The transverse phonon spectra at $\bQ$ = (4, k, 0) across T$_{\rm{CDW}}$ $\approx$ 320 K. Same as Fig.~\ref{figS:FigA_Supplementary_linear} but using a different color scheme, where same colors represent the logarithm of the intensity. (a)-(d) Color maps of the scattering intensity at T~=~300~K, 350~K, 400~K, and 450~K as a function of energy and momentum transfer, k, along the K-direction. The bottom panels, (e)-(h), show the fitted intensity using Eq.~(1) of the main text.}
\label{figS:FigA_Supplementary_log}
\vspace{-0.1in}
\end{figure}

\begin{figure}[h!]
\includegraphics[width=0.5\textwidth]{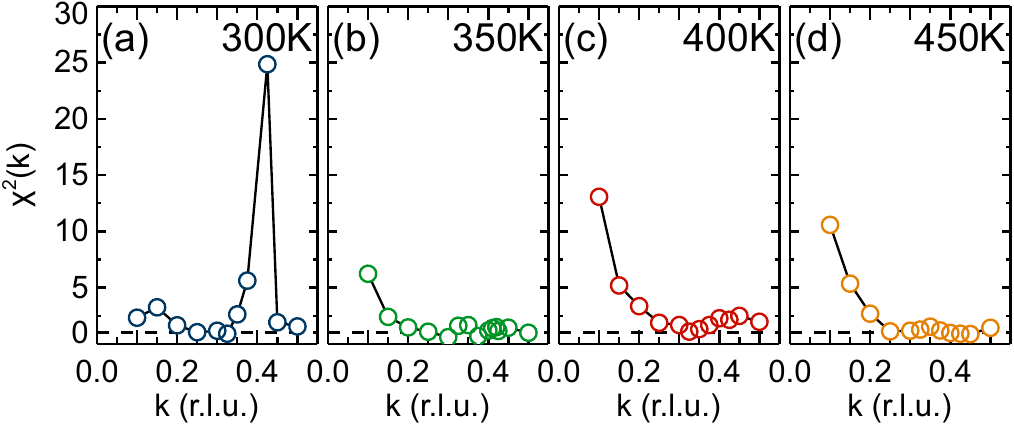}
\caption{(color online) The chi-squared of the fits to the data in Figure~1 of the main text, also shown in Figures~\ref{figS:FigA_new_linear_v1} -\ref{figS:FigA_Supplementary_linear}.}
\label{figS:Fig_chi2_linear}
\vspace{-0.1in}
\end{figure}

Figure~\ref{figS:FigA_Supplementary_linear} presents the comparison of the color maps of the measured intensity (top panels) and the fitted intensity using Eq.~(1) of the main text (bottom panels). Figure~\ref{figS:FigA_Supplementary_log} presents the same data and fits using a different color scheme, where same colors represent the logarithm of the intensity, and which is used in the Figures of the main text. Very good agreement between the data and the fits is observed no matter which color map is used. The good agreement between the fit and the data is quantified by the fit chi-squared, $\chi^2$, presented in Figure~\ref{figS:Fig_chi2_linear}. For most of scans,  $\chi^2 \sim 1$.

\begin{figure}[h!]
\includegraphics[width=0.5\textwidth]{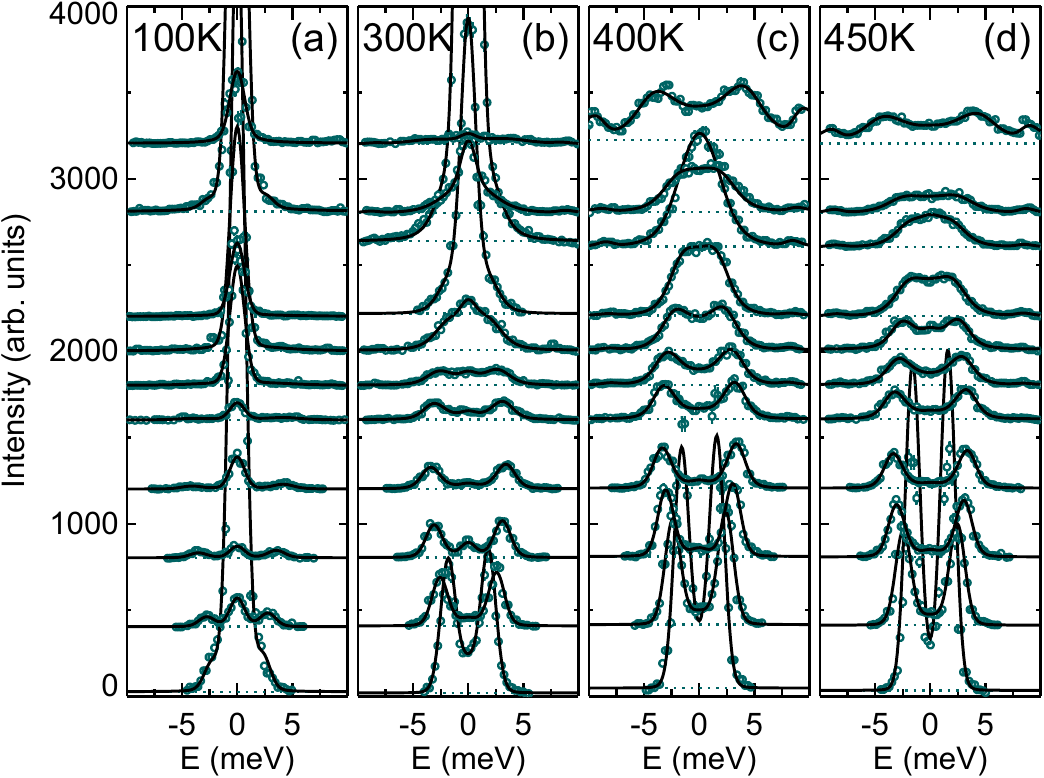}
\caption{(color online) The line scans collected in experiment 06 (symbols) with fits to Eq.~1 of the main text (solid lines). The scans are vertically offset to minimize overlaps; the horizontal dashed lines indicate zero levels. The error bars represent one standard deviation. }
\label{figS:Fig_scans_exp1_linear}
\end{figure}

\begin{figure}[H]
\includegraphics[width=0.5\textwidth]{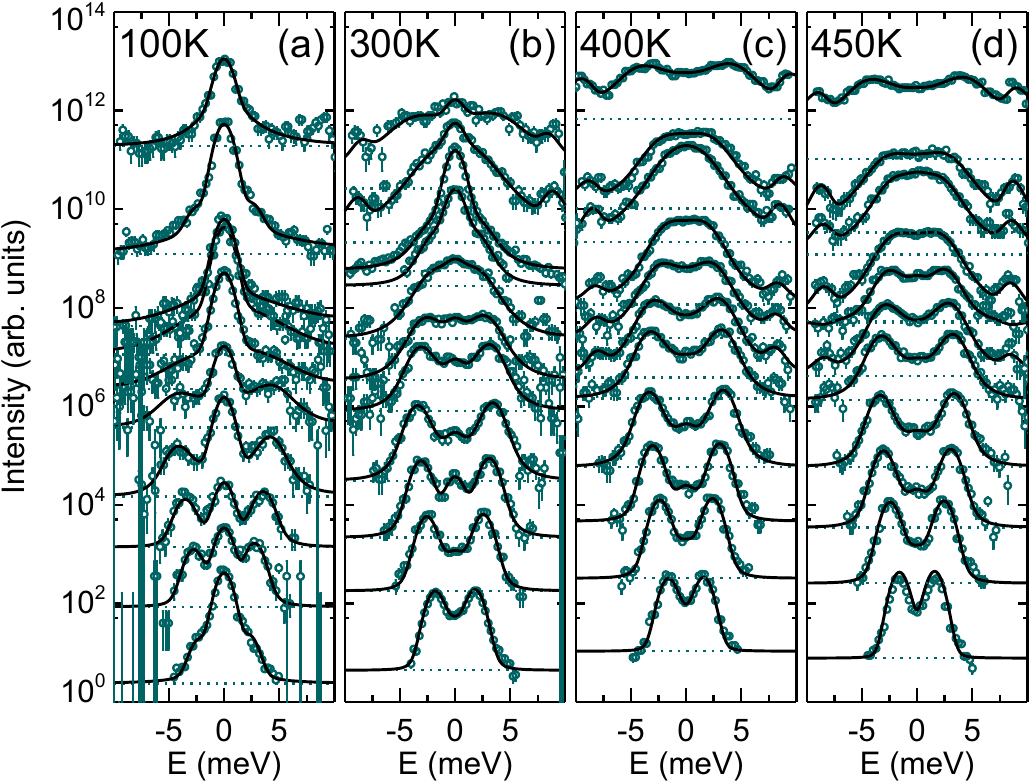}
\caption{(color online) The line scans collected in experiment 06 (symbols) with fits to Eq.~1 of the main text. Same as Fig.~\ref{figS:Fig_scans_exp1_linear} but with the intensity shown on log scale. The scans are vertically offset to minimize overlaps; the horizontal dashed lines indicate zero levels. The error bars represent one standard deviation. }
\label{figS:Fig_scans_exp1_log}
\end{figure}

\begin{figure}[H]
\includegraphics[width=0.5\textwidth]{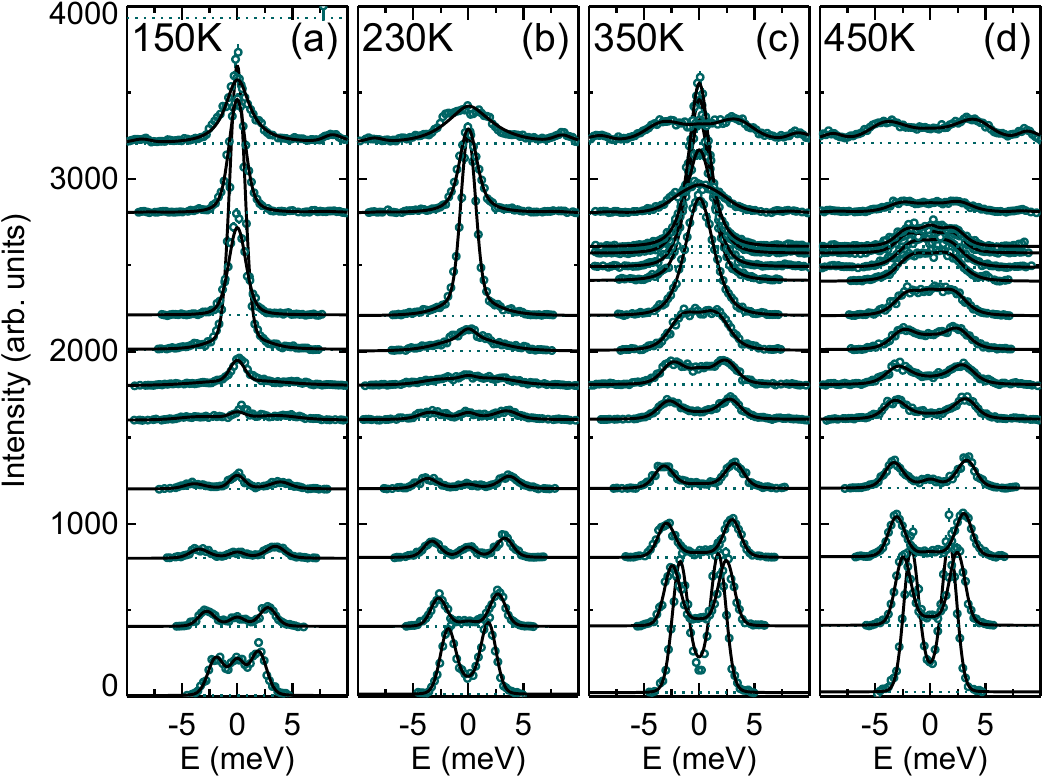}
\caption{(color online) The line scans collected in experiment 12 (symbols) with fits to Eq.~1 of the main text (solid lines). The scans are vertically offset to minimize overlaps; the horizontal dashed lines indicate zero levels. The error bars represent one standard deviation. The scans below T$_{\rm{CDW}}$ (at 150~K and 230~K) in the vicinity of $\rm{\bQ_{CDW}}$ (for k~=~0.41$\pm$~0.015 r.~l.~u.) are off-scale.}
\label{figS:Fig_scans_exp2_linear}
\end{figure}

\begin{figure}[H]
\includegraphics[width=0.5\textwidth]{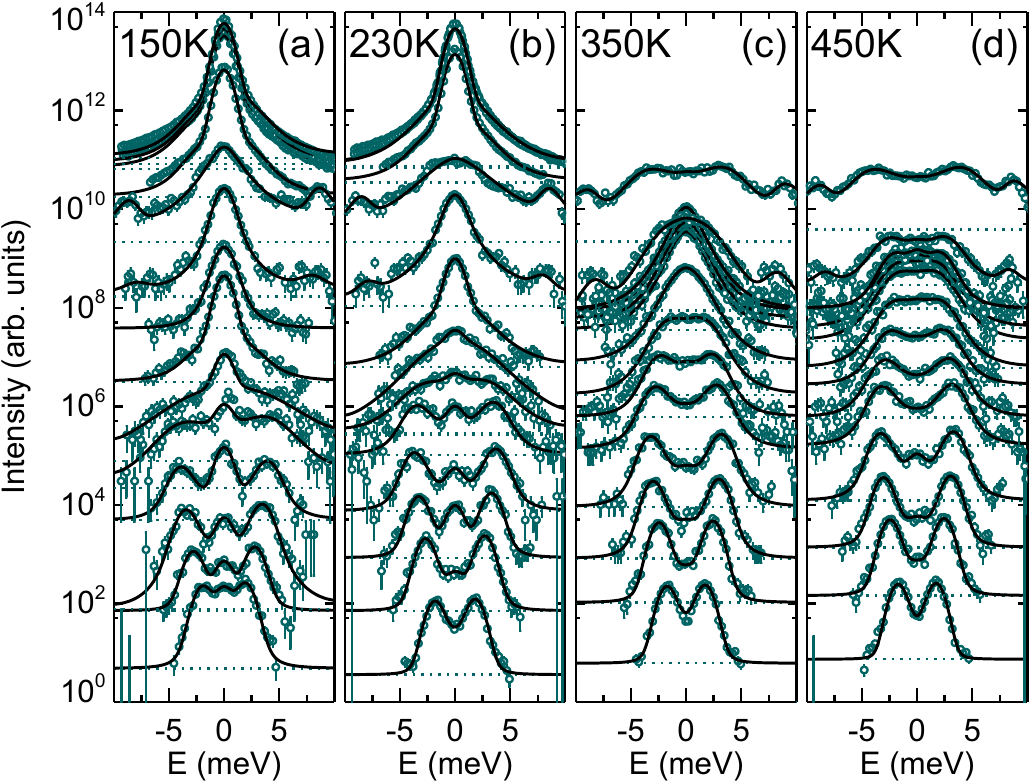}
\caption{(color online) The line scans collected in experiment 12 (symbols) with fits to Eq.~1 of the main text. Same as Fig.~\ref{figS:Fig_scans_exp1_linear} but with the intensity shown on log scale. The scans are vertically offset to minimize overlaps; the horizontal dashed lines indicate zero levels. The error bars represent one standard deviation. }
\label{figS:Fig_scans_exp2_log}
\end{figure}

Finally, for completeness we also include all scans measured in the two experiments, 06 and 12, which we have carried out. Figures~\ref{figS:Fig_scans_exp1_linear} and \ref{figS:Fig_scans_exp1_log} present the data measured in experiment 06 on linear and logarithmic intensity scale, respectively. Similarly, Figures~\ref{figS:Fig_scans_exp2_linear} and \ref{figS:Fig_scans_exp2_log} present the data measured in experiment 06 on linear and logarithmic intensity scale, respectively.

\end{document}